# LA ÚLTIMA FRONTERA DE LA FILOSOFÍA: HACIA UNA SÍNTESIS DE LA ÉTICA DEL FUTURO A LARGO PLAZO, EL RIESGO EXISTENCIAL Y LA ONTOLOGÍA POSTHUMANA

Santos E. Moreta Reyes

*Profesor asociado de Ética en el Instituto Tecnológico de Las Américas, ITLA, RD*

smoreta@itla.edu.do

## Resumen

La humanidad se encuentra en una encrucijada histórica, definida por una capacidad tecnológica sin precedentes y riesgos existenciales concurrentes. Esta coyuntura revela una laguna significativa en la tradición filosófico universal: la ausencia de un marco sistemático y robusto para la filosofía del futuro a largo plazo. Este artículo argumenta que la formulación de dicho marco constituye el imperativo ético y filosófico central de nuestra era. Para defender esta tesis, el análisis procede en cuatro etapas. Primero, se establecen los cimientos normativos del campo, sintetizando el "principio de responsabilidad"de Hans Jonas con la ética consecuencialista impersonal de Derek Parfit. Segundo, se examina el trabajo de Nick Bostrom sobre el riesgo existencial como la aplicación analítica de esta obligación ética, articulando la lógica del largoplacismo (longtermism). Tercero, se aborda el desafío que el posthumanismo y el transhumanismo plantean a la noción de un "sujeto" humano estable, argumentando que cualquier ética futura debe incorporar una ontología fluida del ser. Cuarto, se explora la búsqueda de un propósito cósmico secular como una respuesta funcional al problema de la motivación en un marco no religioso. El artículo concluye que la contribución original de esta síntesis es la articulación de una agenda de investigación para una filosofía prospectiva, una que integre la axiología, la gestión del riesgo y la ontología para guiar a la humanidad a través de su peligrosa adolescencia tecnológica.

**Palabras Clave:** Filosofía del Futuro, Riesgo Existencial, Largoplacismo (Longtermism), Posthumanismo, Ética Intergeneracional, Hans Jonas, Derek Parfit, Nick Bostrom.

## 1. Introducción: La Emergencia de un Nuevo Imperativo Filosófico

Desde sus albores en la Grecia clásica, la filosofía occidental ha centrado sus investigaciones en la condición humana dentro de parámetros ontológicos y temporales que se presuponían estables. La *physis* (naturaleza) constituía un telón de fondo inmutable, y la escala de la acción humana, aunque capaz de producir grandes bienes y catástrofes a escala de una *polis* o incluso de un imperio, no amenazaba la continuidad de la especie ni la integridad de la biosfera a escala geológica. Hoy, esta presunción fundamental ha colapsado. Nos encontramos en el Antropoceno, una nueva época geológica definida por la capacidad humana para alterar sistemas planetarios fundamentales, desde el ciclo del carbono hasta la composición de la estratosfera.[1] Este poder, magnificado por tecnologías de doble uso con un potencial catastrófico sin precedentes —como la inteligencia artificial (IA) general, la biología sintética y el armamento autónomo—, nos sitúa en lo que Toby Ord ha denominado el precipicio.[2]

Esta nueva condición histórica está marcada por una asimetría radical: nuestro poder tecnológico crece de forma exponencial, mientras que nuestra sabiduría colectiva y nuestras instituciones éticas evolucionan de forma lineal y lenta. Esta brecha creciente entre poder y sabiduría no es una mera preocupación teórica; es la fuente de los riesgos más profundos que enfrentamos y exige una reorientación fundamental del quehacer filosófico. Ya no basta con preguntar por la "buena vida en el contexto de una comunidad existente; debemos preguntar cómo asegurar que existan comunidades futuras capaces de hacerse esa misma pregunta.

El presente artículo postula que la construcción de una fi-

---

losofía sistemática del futuro a largo plazo —una filosofía prospectiva— representa la tarea intelectual más urgente y de mayor impacto potencial de nuestro tiempo. Esta no es una mera subdisciplina que pueda añadirse al catálogo filosófico existente, sino una meta-filosofía necesaria para orientar a todas las demás en la era del riesgo existencial. Sostenemos que una empresa de tal magnitud requiere una síntesis ambiciosa e integradora de campos del saber que, con demasiada frecuencia, se han desarrollado de forma aislada.

## 2. Fundamentos Normativos: De la Responsabilidad Asimétrica a la Ética Impersonal

La idea de que tenemos obligaciones morales hacia las generaciones venideras no es completamente novedosa. Se pueden encontrar intuiciones al respecto en el pensamiento de figuras como Edmund Burke, con su noción de la sociedad como un contrato entre los muertos, los vivos y los que están por nacer.[3] Sin embargo, estas formulaciones tradicionales resultan dramáticamente insuficientes para la era tecnológica. Fue el filósofo germano-americano Hans Jonas quien, en su obra magna *El Principio de Responsabilidad* (1984), diagnosticó con mayor agudeza esta insuficiencia.

Jonas argumentó que todas las éticas pre-tecnológicas (desde Aristóteles hasta Kant) compartían tres premisas implícitas: 1) la condición humana era esencialmente fija; 2) el alcance de la acción humana y, por tanto, de la responsabilidad, era limitado en el espacio y el tiempo; y 3) la naturaleza era invulnerable a nuestras intervenciones. La tecnología moderna, según Jonas, ha demolido estas tres premisas. Esto crea una situación ética radicalmente nueva, caracterizada por una asimetría de poder y responsabilidad. Nuestras acciones de hoy tienen consecuencias que se extienden por milenios y afectan a miles de millones de personas que aún no existen. Estas generaciones futuras son radicalmente vulnerables a nuestras decisiones, pero no pueden hacernos responsables. Esta asimetría da lugar a un nuevo imperativo categórico: actúa de tal modo que los efectos de tu acción sean compatibles con la permanencia de una vida humana auténtica en la Tierra".[4] Para guiar la aplicación de este principio, Jonas propone una "heurística del miedo": debemos prestar una atención desproporcionada a los peores escenarios posibles.

Si Jonas estableció la *obligación*, Derek Parfit, en *Razones y Personas* (1984), nos proporcionó las herramientas para pensar en esa obligación con rigor lógico. Parfit introdujo el famoso "Problema de la no Identidad".[5] El problema sur- ge al considerar políticas a gran escala que afectan a quie-nes nacerán. Imaginemos dos políticas energéticas: una de "business-as-usual" que causa una catástrofe climática en 200 años, y una "sostenible" que la evita. Debido a sus efectos, las personas que nacen bajo cada política son completamente diferentes. La paradoja es que no podemos decir que la primera política "daña a las personas del futuro, porque esas personas específicas no habrían existido de otra manera.

La solución radical de Parfit, es un giro hacia una ética impersonal y consecuencialista. Lo que importa moralmente no es si dañamos a individuos específicos, sino si el *resultado* es peor. Un mundo con una catástrofe climática es peor que un mundo sin ella, independientemente de la identidad de sus habitantes. Desde esta perspectiva, la extinción es la peor de las catástrofes porque aniquila la totalidad del valor futuro potencial.[6]

Juntos, Jonas y Parfit nos proporcionan el andamiaje normativo necesario. Jonas nos da el "porqué": una obligación asimétrica de asegurar la continuidad de una humanidad futura. Parfit nos da el cómo pensar sobre ello, de una manera impersonal y enfocada en el valor total.

## 3. La Formalización del Deber: Riesgo Existencial y la Lógica del Largoplacismo

Sobre estos cimientos, Nick Bostrom ha erigido un programa de investigación centrado en el concepto de riesgo existencial: aquel que amenaza con aniquilar la vida inteligente originaria de la Tierra o restringir permanente y drásticamente su potencial.[7] Esta definición lo distingue de otras catástrofes, pues su daño es terminal e irrecuperable.

Los riesgos se clasifican en naturales (asteroides, supervolcanes) y antropogénicos. Estos últimos son la principal preocupación e incluyen la guerra nuclear, pandemias sintéticas o una inteligencia artificial no alineada.[8]

Esta formalización da lugar al largoplacismo (*longtermism*), la visión de que influir positivamente en el futuro a largo plazo es una prioridad moral clave.[9] Su argumento se basa en el cálculo del valor esperado: si la humanidad sobrevive, el número de vidas futuras potenciales es astronómicamente grande.[10] Por tanto, incluso una pequeña reducción en la probabilidad de un riesgo existencial produce un beneficio moral de magnitud extraordinaria. El largoplacismo traduce

---

así el imperativo abstracto de Jonas y Parfit en un programa de acción concreto y priorizado.

# 4. El Problema del Sujeto: Posthumanismo y la Ontología del Futuro

Una filosofía del futuro debe responder: ¿el futuro de *quién* o de *qué* estamos tratando de asegurar? La concepción de "humanidad como categoría estable es cuestionada por el posthumanismo y el transhumanismo.

El posthumanismo crítico (Rosi Braidotti, Francesca Ferrando) deconstruye el humanismo como una ideología antropocéntrica.[11] Aboga por una visión donde se difuminan las fronteras entre lo humano, lo animal y lo tecnológico. Esto desafía a la ética largoplacista a aclarar si su objetivo es la preservación de *Homo sapiens* o la propagación de la conciencia y la complejidad, independientemente de su sustrato. Por otro lado, el transhumanismo (Nick Bostrom, Ray Kurzweil) aboga activamente por la superación de las limitaciones humanas mediante la tecnología, planteando la posibilidad de sucesores "posthumanos con capacidades radicalmente mejoradas.[12] Esto introduce la cuestión de si tenemos la obligación de permanecer "humanos o de guiar nuestra propia evolución.

Una filosofía del futuro a largo plazo, por tanto, no puede ser una mera ética de la *preservación*; debe ser también una **ontología política** que delibere sobre la naturaleza deseable del sujeto moral del futuro.

# 5. Discusión y Limitaciones del Enfoque

La síntesis propuesta no está exenta de importantes desafíos. A continuación, los tres más importantes:

## 5.1. Incertidumbre Epistémica Radical

La crítica más persistente es el problema de la ignorancia o *cluelessness*.[13] ¿Cómo podemos conocer los efectos a muy largo plazo de nuestras acciones? La defensa no es afirmar que poseemos una ciencia predictiva, sino que operamos desde un marco de gestión prudente del riesgo bajo incertidumbre, priorizando estrategias robustamente buenas (ejemplo: mayor cooperación internacional).

## 5.2. La Tensión entre Presente y Futuro

Existe una tensión normativa entre las obligaciones hacia el futuro lejano y las demandas urgentes del presente (pobreza, enfermedades).[14] ¿Es moral desviar recursos de salvar una vida hoy para mitigar un riesgo futuro? Los largoplacistas argumentan que, por la magnitud de lo que está en juego, la mitigación del riesgo existencial sigue siendo prioritaria, y que muchas intervenciones benefician también al presente.

## 5.3. El Problema de la Motivación

En un mundo secular, ¿qué narrativa puede movilizar la acción colectiva a la escala requerida? La idea de un "propósito cósmico" secular, explorada por filósofos como Philip Goff, puede tener un valor funcional.[15] La visión de la humanidad como portadora de la conciencia en un universo mayormente vacío podría proporcionar la inspiración necesaria para sostener un esfuerzo intergeneracional.

# 6. Conclusión y Agenda de Investigación Futura

Este artículo ha argumentado que la filosofía debe desarrollar un marco sistemático para el futuro a largo plazo, sintetizando cuatro dominios: 1) una base normativa (Jonas, Parfit); 2) una metodología analítica (Bostrom); 3) una ontología crítica del sujeto posthumano; y 4) una teleología funcional para la motivación.

La contribución de este análisis es su *articulación sintética como un programa de investigación unificado*. La agenda futura debe ser interdisciplinaria, con líneas prioritarias como:

1. **Justicia Intergeneracional y Global:** Desarrollar teorías que equilibren las demandas del presente y el futuro.

2. **Axiología del Riesgo y del Futuro:** Refinar la teoría del valor para tomar decisiones bajo incertidumbre radical.

3. **Ética de la Creación y la Transición:** Elaborar una ética normativa para la era posthumana (IA, genética).

4. **Psicología y Sociología de la Orientación a Largo Plazo:** Investigar cómo traducir los imperativos filosóficos en acción efectiva.

En última instancia, forjar la brújula conceptual para navegar nuestro futuro no es solo una tarea académica; es una condición necesaria para que la vasta y prometedora historia de la humanidad tenga la oportunidad de ser escrita.

---

[11]Ver Braidotti, R. (2013). *The Posthuman*. Polity Press; y Ferrando, F. (2019). *Philosophical Posthumanism*. Bloomsbury Academic.

[12]Bostrom explora esta idea en numerosos trabajos. La analogía busca ilustrar que las capacidades posthumanas podrían estar en dimensiones que nos son, por ahora, inconcebibles.

[13]El término fue acuñado por el filósofo James Lenman. Ver Lenman, J. (2009). Cluelessness. *The Proceedings of the Aristotelian Society, Supplementary Volumes*, 83, 1–20.

[14]Esta crítica es a menudo resumida bajo el lema "los problemas del presente primero» es un punto central de debate dentro del movimiento del altruismo eficaz.

[15]Goff, P. (2023). *Why? The Purpose of the Universe*. Oxford University Press. Goff argumenta desde una perspectiva panpsiquista que el universo podría tener un propósito inherente que la vida inteligente está destinada a cumplir.

# Referencias Bibliográficas